\documentstyle[epsfig]{mn}
\def\be{\begin{equation}}
\def\ee{\end{equation}}
\def\bea{\begin{eqnarray}}
\def\eea{\end{eqnarray}}
\def\etal{{et al.}\thinspace}

\begin{document}

\title
[Heating of the intracluster medium by quasar outflows
]
{Heating of the intracluster medium by quasar outflows}

\author[Biman B. Nath and Suparna Roychowdhury]
{Biman B. Nath and Suparna Roychowdhury\\
Raman Research Institute, Bangalore 560080, India\\
(biman@rri.res.in, suparna@rri.res.in)
}
\maketitle

\begin{abstract}
{
We study the possibility of quasar outflows in clusters
and groups of galaxies heating the intracluster gas in order to explain 
the recent observation of excess entropy in this gas. We use the extended
Press-Schechter formalism to estimate the number of quasars that
become members of a group of cluster of a given mass and formation epoch.
We also estimate the fraction of mechanical energy in the outflows
that is imparted to the surrounding medium as a function of the
density and temperature of this gas. We finally calculate the total
amount of non-gravitational heating from such outflows 
as a function of the
cluster potential and formation epoch. 
We show that outflows from broad absorption line (BAL) and radio loud quasars
can provide the required amount of heating of the intracluster gas. 
We find that in this scenario
most of the heating takes place at $z \sim 1\hbox{--}4$, and that this
``preheating'' epoch is at lower redshift for lower mass clusters.
}
\end{abstract}

\begin{keywords} Cosmology: Theory---Galaxies: Intergalactic Medium---
X-rays: Galaxies: Clusters
\end{keywords}

\section{Introduction}

Clusters and groups of galaxies contain a large amount of hot gas, besides
galaxies and the gravitationally dominant dark matter. This hot X-ray-emitting 
gas known as the intracluster medium (ICM) represents a part of the
baryonic matter of the universe that is not associated with individual galaxies
but remains trapped in the deeper gravitational potential of galaxy clusters.
Hierarchical models of structure formation have been very successful in
explaining many observed properties of galaxies and galaxy clusters.
Nevertheless, some puzzling problems remain open and unexplained. Models
of cluster formation in which the intergalactic gas simply falls into the dark
matter  dominated gravitational potential well (so-called infall models) fail to
reproduce all the structural properties of the local cluster population
(e.g., Evrard \& Henry 1991; Navarro, Frenk \& White 1995; Mohr \& Evrard
1997;
Bryan \& Norman 1998). There is certainly some additional physics driving ICM
evolution.

Recent X-ray observations have provided evidences for 
some non-gravitational
heating of the diffuse, high density baryons in the potential
wells of groups and clusters of galaxies, in addition to the heating during
the gravitational collapse. One of the first evidences was in the shape 
of the $L_x-T$
relation, which is steeper than the self-similar behaviour $L_x \propto T^2$
predicted in the case of gravitational processes only. As early as the
emergence of $ROSAT$ and $Einstein$ data, several authors proposed
that the missing element is the existence of a ``preheated high-entropy"
intergalactic gas prior to a cluster's collapse (David \etal 1991; Evrard
\& Henry 1991; Kaiser 1991; White 1991). Later, Ponman \etal (1999), and
Lloyd-Davies \etal (2000)
found direct evidence of an entropy excess with respect to 
the level expected from
gravitational heating in the centre of groups. Ignoring the constant
and logarithms, one can define the ``entropy'' as
$S\equiv T/n^{2/3}$. The excess entropy, or equivalently, the excess specific
energy, flattens the density profile decreasing the X-ray luminosity which
is proportional to the square of the density. The effect is stronger in
poorer clusters, where the excess energy associated with the excess
entropy is comparable
to the gravitational binding energy, while rich clusters, 
where gravity is dominant,
are mostly unaffected. This produces a steepening of the $L_x-T$ relation.

The most popular scenario to successfully explain these thermal
properties of the ICM has been the ``preheating" scenario. For this scenario,
the candidate processes which have been looked into are  strong galactic
winds driven by supernovae. However Valgeas \& Silk (1999) showed that the
energy provided by supernovae cannot raise the entropy of intergalactic
medium (IGM) up to the level required by current observations. The observed
amount of required energy injection depends on the epoch, and have been
in the range of 0.4 - 3 KeV per gas particle (Navarro et al. 1995;
Cavaliere et al. 1997; Balogh et al.1999; Wu et al. 2000; Lloyd-Davies \etal 
2000, Borgani \etal 2001). For example,
Wu \etal (2000) showed that galactic winds can impart only $\le 0.1$ keV
per particle. Moreover,
Kravtsov \& Yepes (2000) estimated the energy provided by supernovae from
the observed metal abundance of ICM and found that the heating only by
supernovae driven outflows requires unrealistically high efficiency. 
On the other hand, quasar outflows
 may be much more powerful and plausible candidates of the heating
(Valageas \& Silk 1999). In this paper, we focus on the role of quasar outflows
in this regard.

The epoch of the energy input also remains uncertain.
Lloyd-Davies \etal (2000) put an upper limit of 
$z_{max} \sim 7\hbox{--}10$ on the preheating epoch, from their estimate of
excess entropy in groups. 
For AGNs, there have been
no additional constraints like the metal abundance in the case of supernova
heating (Kaiser \& Alexander 1999). Recently, 
Yamada \& Fujita (2001) have looked into the deformation of the
cosmic microwave background (the Sunyaev-Zel'dovich effect) by hot electrons
produced at the shocks produced by jets from AGNs. They showed that the
observed excess entropy of ICM and $COBE/FIRAS$ upper limit for Compton
$y$-parameter are compatible with each other only when the heating by jets
occurred at relatively small redshift ($z\leq 3$). Thus they questioned the
``preheating" scenario as their result suggests that the heating occurred
simultaneously with or after cluster formation.

In this paper, we calculate the heat input from quasar outflows inside
clusters. We calculate the mechanical work done by various kinds of 
quasar outflows and the excess
energy imparted by them onto the intracluster medium via pdV work.
For the statistics of quasars inside
clusters, we use the extended Press-Schechter formalism. Finally we
calculate the excess energy per particle 
and tally them with available observations.

In the next section, we discuss the abundance of quasars inside
clusters of a given mass. We then discuss the evolution of quasar outflows
and the mechanical work done by them in \S 3. We use these
concepts to calculate the heating
of the ICM in \S 4. We then discuss the implications of our results in \S 5.
Throughout the paper we assume a flat universe with a cosmological constant,
with $\Omega_0=0.35, \, \Omega_{\Lambda 0}=0.65,$ and $ h=0.65$.

\section{Quasars inside clusters}

For a proper evaluation of the heat input from quasars inside
clusters, one first needs to calculate their abundance and its dependence
on the quasar mass, cluster mass, and the cluster formation redshift.
Observationally, it is still difficult to obtain good statistics of quasars
inside clusters. Estimation of the galaxy-QSO correlation function
have shown that at low redshifts ($z \la 0.4$), quasars typically reside
in small to moderate groups of galaxies and not in rich galaxies (e.g.,
Bahcall \& Chokshi 1991; Fisher \etal 1996). It is still uncertain whether
or not their optical or radio luminosities depend on their environments. 
Some studies (e.g., Ellingson, Yee \& Green 1991) have shown that
radio-loud quasars preferentially reside in richer clusters at higher redshift,
although more recent studies (e.g., Wold \etal 2001) do not find any
such correlation.

There has been, however, considerable work in relating the observed
quasar luminosity function or the radio luminosity function (for radio-loud
quasars) with the mass function of galaxies as prescribed by
the Press-Schechter (PS) formalism (e.g., Haehnelt \& Rees 1993; Haiman
\& Loeb 1998; Yamada, Sugiyama \& Silk 1999).

At $z=0$, Yamada, Sugiyama \& Silk found that one can reproduce the
abundance of the radio sources powered by AGNs (i.e., leaving aside
the radio sources powered by starbursts) by assuming that a fraction
$f_r$ of the halos from PS formalism become radio-loud quasars, where,
$f_r \sim 0.01$ for $M_h \ga 10^{12}$ M$_{\odot}$, and $f_r=0$
for $M < 10^{12}$ M$_{\odot}$. They assume an upper limit on $M_h$
of $10^{14}$  M$_{\odot}$. Since it is known that radio loud quasars
constitute a fraction $0.1$ of the quasar population (Stern \etal 2000), 
this means that
a fraction $f_q=10 f_r$ of the halos from PS formalism become quasars.
In other words, $f_q \sim 0.1$ for $10^{12} \le M_h \le 10^{14}$ 
M$_{\odot}$, and $f_q=0$, otherwise.
Yamada, Sugiyama \& Silk (1999) assumed this fraction to be a constant for
all redshifts.
In this model, the rate of formation of quasars is given by
the derivative of the PS mass function at the relevant mass scale.

This is similar to the model adopted by Haiman \& Loeb (1998) and
Furlanetto \& Loeb (2001; FL01). FL01 showed that at 
high redshift ($z \ga 4$) the
rate of formation of quasars is a fraction $f_q \sim 0.1$ of the
rate of formation of halos from PS formalism, if a life time
of order $10^7$ yr is assumed for the quasar. 
There is, however,
a difference, in that they had only a lower limit to $M_h$ and
no upper limits (see eqns (2) \& (3) of Haiman \& Loeb 1998). They
mention that at low redshifts their formalism does not predict
any decline as observed in reality, and for which they consider their
model only at high redshift. It is possible that this mismatch is
due to the lack of upper limits, since the 
the differentiation of PS mass function for objects 
with an upper limit in mass decreases at low redshift,
(Haiman, Z. 2001, private communications). In any case, 
Haiman \& Loeb (1998) found
that this prescription yields a matching quasar luminosity function
that is observed, at redshift $z \ge 2.5$. We find later that most 
of the heating of the ICM gas (even for our least massive cluster)
occurs at $z \ge 2$ (Figure 7). We will therefore assume for simplicity that 
this fraction $f_q \sim 0.1$ at all redshifts (as in Yamada \etal 1999).

In this paper, we would like to have a conservative estimate of the quasar
abundance. Also, we would like to calculate the abundance of quasars in
clusters including low mass groups of galaxies.
For this reason, we assume
the value of $f_q$ as above, but use an upper limit of $10^{13}$ 
M$_{\odot}$. Since the mass function decreases steeply at the higher mass
end, this should not change the value of $f_q$ substantially.
In brief, we assume that,
\be
f_q \sim \left\{ \begin{array}{ll}
		0.1 & \mbox{if $10^{13} \ga M_h \ga 10^{12} \, M_{\odot}$} \\
		0  &	\mbox{if $M_h > 10^{13} \, M_{\odot} \,,
M_h < 10^{12} \, M_{\odot}  \> ,$}
		\end{array}
	\right.
\label{eq:fq}
\ee
motivated by the
model of Yamada, Sugiyama \& Silk (1999), and by the fact that a similar
prescription by Haiman \& Loeb (1998) recovers the quasar population at
high redshift, and relate the rate of 
formation of quasars with that of halos in the PS formalism.

We are, however, concerned with the statistics of  quasars
{\it inside} clusters. For this one needs to have an extension of the
PS mass function which can predict the probability of a given halo
becoming a part of bigger object later, or the probability of an object
having had a progenitor of a given mass at an earlier epoch. Such
extensions of the PS theory have been studied in detail by Bower (1991)
and Lacey \& Cole (1993), for example. 

In the standard PS theory, the mass function, i.e., the fraction of 
regions with mass in the range $M, M+dM$ 
which have overdensity $\delta$ in excess of
$\delta_c$ (which is the threshold for perturbations becoming
non-linear), is given by,
\be
f_{PS} = f(M,\delta_c) dM={-1 \over \sqrt{2 \pi}} {\delta_c \over (\sigma_M^2)^{3/2}}
\exp \Bigl [ -{\delta_c^2 \over 2 \sigma_M^2} \Bigr ] {d \sigma_M^2 \over dM}
dM \,.
\label{eq:ps1}
\ee
Here, $\sigma_M$ is the mass variance of the perturbation at the mass scale 
$M$. The relation between the number density of objects in the mass range
$M, M+dM$ with $f(M)dM$ is,
\be
n(M)dM={\rho_o \over M} f(M) dM \,,
\label{eq:ps2}
\ee
where $\rho_o$ denotes the background mass density.

In the extended PS theory, the fraction of regions of mass $M$, contained
within a larger scale region of mass $M'$ and overdensity $\delta '$,
which are more overdense than $\delta_c$, is given by,
\bea
f(M,\delta_c \vert M', \delta ') dM &&=
{-1 \over \sqrt{2 \pi}} {(\delta_c - \delta ') \over (\sigma_M^2 -
\sigma_{M'}^2)^{3/2}} \nonumber\\ && \times
\exp \Bigl [ -{(\delta_c -\delta ')^2 \over 2 (\sigma_M^2 -\sigma_{M'}^2)} 
\Bigr ] {d \sigma_M^2 \over dM}
dM \,.
\label{eq:ps3}
\eea
This expression recovers the simple PS mass function in the limit
$M' \rightarrow \infty$ and $\delta ' \rightarrow 0$, relevant for the
whole universe.

If we then identify $M'$ with $M_{cl}$, the mass of a cluster, and
$\delta '=\delta_c(z_f)$, the threshold overdensity of the cluster at
its formation epoch $z_f$, we can then obtain the mass fraction of $M_{cl}$
which have been parts of progenitors of a given mass range ($M$ to $M+dM$)
at a given (earlier) redshift. If we also identify this mass range with that
of the quasars as in the standard PS theory ($10^{12} \hbox{--} 10^{13}$
M$_{\odot}$), and use the fraction of these halos that become quasars
($f_q$; eqn \ref{eq:fq}), 
we will obtain the mass fraction of the final cluster which
have been quasars at some given earlier epoch.

To obtain the rate of formation of these quasars (inside a future
cluster of mass $M_{cl}$), we should differentiate the above expression.
A simple differentiation  will, however, not give the correct result,
since there will be a negative contribution from the merging of halos
out of this mass range. One would get a negative rate of formation
of such quasars  at some point
if the rate at which they disappear beyond this mass limit
is not taken into account in a proper manner. This 

Consider the abundance
of objects in a given mass range $M, M+dM$ at two successive epochs
$z_1$ and $z_2$ ($z_2 < z_1$). The abundance $f(M)dM$ at $z_2$ will
be given by $f_2=f_1 + F -D$, where $f_2$ and $f_1$ are the abundances
at epochs $z_2$ and $z_1$, the term $F$ denotes the abundance of newly
formed objects in this mass range during the epoch $z_1$ and $z_2$
(from merger of smaller objects),
and $D$ signifies the abundance of objects that moved out of this mass range
as a result of merger (into bigger objects). A simple differentiation of the
PS function, involving the difference $(f_2-f_1)$ will therefore depend
on both $F$ and $D$. For a given range of mass, $D$ is very small
at a very early epoch, but it 
increases with time (see, e.g., Fig. 5 of Haiman \& Menou 2000),
and at a later epoch can become larger than $F$.
Therefore, at lower redshifts, a simple differentiation can imply
a negative rate of change of abundance. If the contribution of $D$
is neglected, one would then incorrectly get a negative value of $F$.

This problem has been encountered in the case
of ordinary PS function by many authors. While studying the rate of
mergers in the context of background radiation from starbursts and AGNs,
Blain \& Longair (1993) noted that a simple differentiation of the PS
function leads to a negative rate of formation of objects in a given
mass range. In other words, the actual rate of formation of objects is
given by,
\be 
\dot f_{form}=\dot f_{PS} + \dot f_{merger}  \,.
\label{eq:merg1}
\ee
They performed a simulation assuming a simple power spectrum and
obtained a fit for the rate of these objects merging to form bigger objects.
They found that $\dot f_{merger}$ can be approximated well (in the
Einstein de-Sitter universe) by,
\be
\dot f_{merger}= \phi {f_{PS} \over t} \exp \Bigl [ (1-\alpha) {\delta_c^2
\over 2 \sigma^2} \Bigr ] \,
\ee
where the value of $\alpha \sim 1.35$ and $\phi \sim 1.3 \hbox{--} 1.7$. 
This problem has also been investigated by Sasaki (1994) and Percival
\& Miller (1999). With the extension of the PS formalism, one can now 
calculate this merging rate (see also, Chiu \& Ostriker 2000).

We note here that in the case of the merger of a lower mass quasar
into a more massive quasar, our implicit assumption is that the central
black holes also merge and form a bigger black hole appropriate for the
bigger quasar (see below; eqn \ref{eq:bh}). 
In other words, we assume that the central
black hole mass always traces the halo mass. 

The rate at which an object of mass $M$ at epoch $z$ merges to form
a bigger object of mass $M'$ is given by (Bower 1991; Lacey \& Cole 1993),
\bea
{d^2 p \over dM' dz} (M \rightarrow M' \vert z) dM' &=&
{1 \over \sqrt{2 \pi}} {\sigma_M^2 \over \sigma_{M'}^2(\sigma_M^2 -
\sigma_{M'}^2)^{3/2}} \nonumber\\ && \times
\exp \Bigl [ -{\delta_c(z)^2 (\sigma_M^2 - \sigma_{M'}^2) 
\over 2 \sigma_M^2 \sigma_{M'}^2)}
\Bigr ] \nonumber\\
&& \times 
\vert{d \sigma_M^2 \over dM}\vert \vert {d \delta_c(z) \over dz} \vert
dM \,.
\label{eq:lc}
\eea
Here, $p (M \rightarrow M' \vert z) dM'$ is the probability of an
object of mass $M$ merging to become an object of mass within the range
$M', M'+dM'$ at redshift $z$.
The rate of disappearance of objects of a given mass, $\dot n_{merger} $,
should be essentially,
\be
{d f_{merger} \over dz} (M,z)= f_{PS} (M,z) 
\int_{2 M}^\infty {d^2 p \over dM' dz} 
(M \rightarrow M' \vert z) dM' \,,
\label{eq:int}
\ee
where the lower limit of the integration is chosen to be such that the
merged object is at least twice as massive as the merging object. We
show the result of this integration for a sCDM universe (with a COBE
normalized power spectrum) in Figure 1, and show the fit of Blain
\& Longair with $\phi=0.9$ and $\alpha=1.35$. We find that the merging
rate is fit by a lower value of $\phi$ than they assumed, although the
difference is a factor of order unity. It is possible that this difference
is due to the specific assumption in the simulation done by Blain
\& Longair (1993), e.g., in the power spectrum being a simple power law
(Blain, A. 2001, private communication), 
or it can be a result of the lower limit
($2M$) chosen by us. At any rate, if we chose the above integral to 
represent the merger rate then it would be a conservative estimate,
since decreasing the lower limit would simple increase the value of the
merger rate, and in turn, the formation rate of objects. 

\begin{figure}
\centerline{
\epsfxsize=0.6\textwidth
\epsfbox{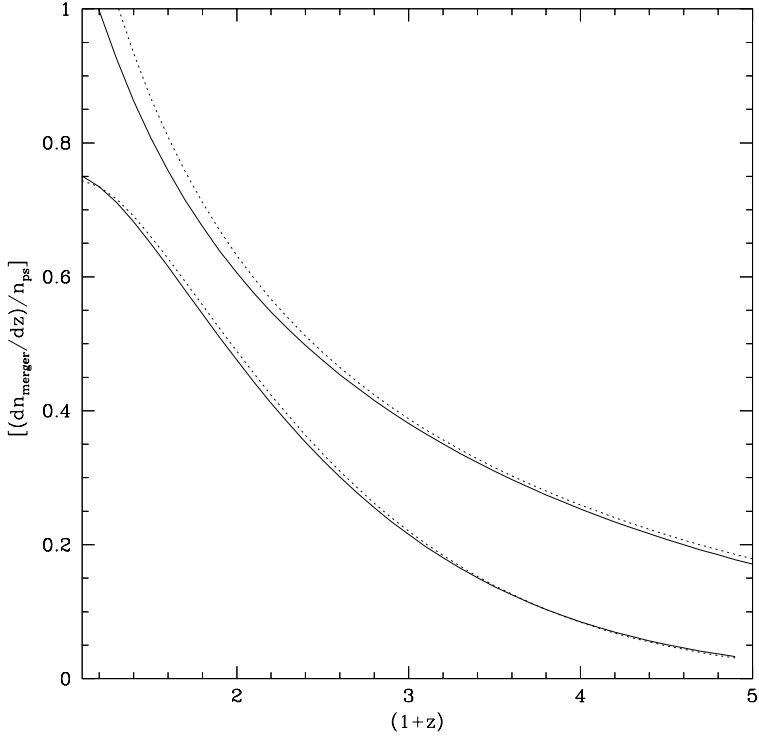}
}
{\vskip-3mm}
\caption{
The rate of disappearance of objects is compared  in different
cosmologies. The thin solid line shows the result of integration
in eqn (\ref{eq:int}) for a SCDM universe for $M=10^{14} \, M_{\odot}$
and the dotted line shows the Blain \& Longair
(1993) fit with $\phi=0.9$ and $\alpha=1.35$. The thick solid line
shows the result of the integration in the $\Lambda$CDM universe
and the dashed line refers to a fit described in the text. Both
curves use 4-year COBE normalized spectra.
}
\end{figure}

We have found that
the result in the case of the $\Lambda$CDM universe can be fit by a similar
function, with $\phi$ being replaced by $0.9 \,{d \delta_c(z) \over dz}$, for
$\alpha=1.35$
with an accuracy of order $\la 5 \%$. We, however, do not
use these fits in our calculation, and evaluate the integral numerically
for our purpose.

To be precise,
this rate of disappearance is valid for the objects following the
PS mass function, i.e., for objects which are not already parts of bigger
objects. Motivated by the extension of the PS formalism, we here posit
that the rate of disappearance of objects inside a bigger object also
has the same form, with $f_{PS}(M,\delta_c)$ in eqn (\ref{eq:int}) 
being replaced by 
$f(M,\delta_c \vert M', \delta ')$. There is admittedly no way of
verifying the truth of this {\it ansatz} at present, since this would involve
more extensions of the PS theory than that is available now. It will also
involve comparing the merger rates inside and outside of clusters. It, however,
leads to a conservative estimate for the formation rate of quasar in a cluster.
As the work of Bower (1991) has shown, growth of perturbations
 {\it inside} a cluster
is enhanced compared to in the field. This means that the merging rate of
objects of a given mass inside a cluster should be larger than that in
the field. Here, by assuming a comparable merging probability (the factor
that multiplies the abundance of objects, given by $f_{PS}$), we are
in a way underestimating the rate of disappearance ($\dot f_{merger}$), 
and in turn, the
rate of formation ($\dot f_{form}$) of quasars in a cluster. The final result
of total heat input from our formalism should, therefore, be a conservative
estimate.

\begin{figure*}
\begin{minipage}{140mm}
\begin{center}
{\vskip-4mm}
\psfig{file=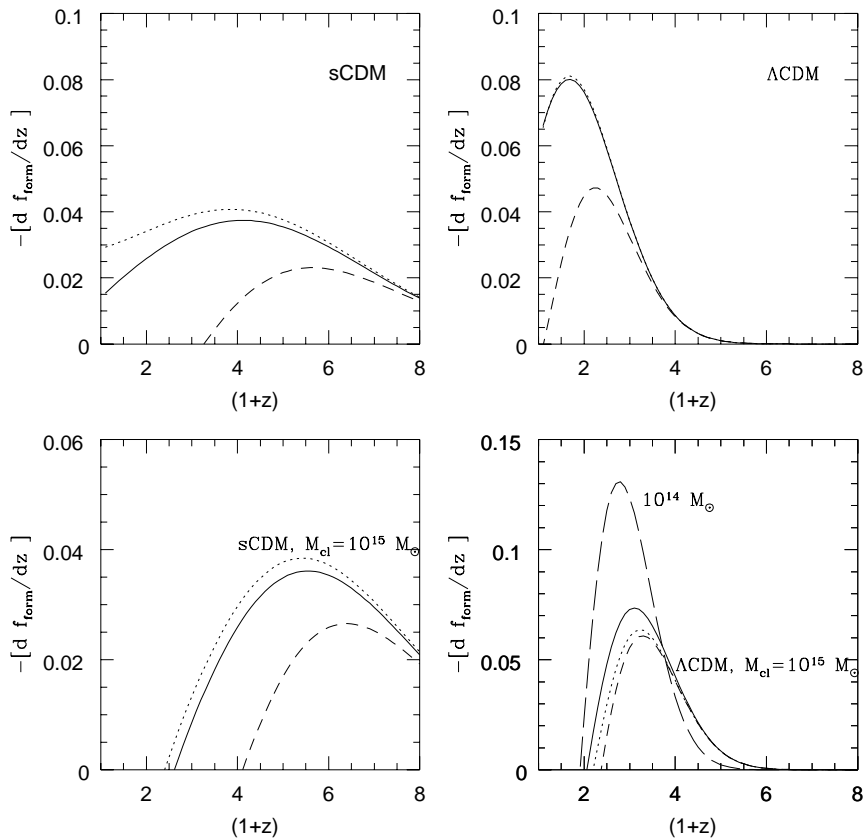, width=12cm}
\end{center}
{\vskip-3mm}
\caption{
The net formation rate of objects (${d f_{form} \over dz}$) is plotted for
objects of mass $10^{13}$ M$_{\odot}$ (for $dM=M$), in clusters
(bottom panels, with $M_{cl}=10^{15}$ M$_{\odot}$) and in general (top
panels), for sCDM (left panels) and $\Lambda$CDM (right panels) models.
Dashed lines show the term ${d f_{PS} \over dz}$, dotted lines show
${d f_{form} \over dz}$ using the Blain \& Longair fit (with $\phi=1.0$),
and solid lines show  ${d f_{form} \over dz}$ using the integral in
equation \ref{eq:int}. The long-dashed line in lower right panel shows
the case for $M_{cl}=10^{14}$ M$_{\odot}$.
}
\end{minipage}
\end{figure*}

We show the results of adding the rate of disappearance in Figure 2. The
dashed lines, which show the term $-(d f_{PS} /dz)$ (or equivalenlty,
${d f_{PS} \over dt}$), become negative at
lower redshift, 
suggesting the need for the addition of the rate
of disappearance of objects. The dotted lines show the result of adding
this rate, using the  Blain \& Longair (1993) fit  (with $\phi=0.9$),
and the solid lines show the result using the integral in equation
\ref{eq:int}. The upper panels of the figure show the case of objects
(of mass $10^{13}$ M$_{\odot}$) in the field, and the bottom panels show
the case for these objects inside a cluster of $M_{cl}=10^{15}$ M$_{\odot}$
($z_f=0$). For these curves, we have used the integration in equation
\ref{eq:int}, and the extension of the PS mass function for the abundance
of objects inside the cluster, as explained above. 
The left and right panels show the cases for sCDM and
$\Lambda$-CDM universe. As the bottom panels show, the addition of the
disappearance rate does not suffice to make $[-(d f_{PS} /dz)]$ (inside
clusters) positive
at low redshift. This is suggestive of the enhanced growth of perturbation,
and the need for a {\it larger} rate of merger {\it inside} clusters,
as previous authors have noted.

We also show the case for $M_{cl}=10^{14}$ M$_{\odot}$ by the long dashed
line in the lower right panel of Figure 2. Comparison with the solid line
(for $M_{cl}=10^{15}$ M$_{\odot}$) shows that the formation rate of these
galaxies inside a lower massive cluster is larger. This follows simply
from the extension of the PS formalism (from the dependence on the
term $(\sigma_M^2-\sigma_{M'}^2)$). It is interesting to note that this
is consistent with the suggestion from observation that quasars are
preferentially located in groups of galaxies instead of rich clusters
(e.g., Bahcall \& Chokshi 1991; Fisher \etal 1996).

If our formalism is used without any correction, this will lead to 
{\it subtraction} of energy input in the final result.
We circumvent this problem by putting $[-(d f_{PS} /dz)]=0$ when this
term turns negative. This will, therefore, provide a lower limit to the
total energy input from quasar outflows in a cluster.

We can finally write down the rate of formation of quasars in a
given mass range $M, M+dM$ 
inside a cluster of a given mass, $M_{cl}$, in the form of the rate
of increase of the fraction of mass of $M_{cl}$ which is in quasars
at an epoch $z$, as (remembering the equation \ref{eq:ps2}),
\bea
{d f_{q,cl} (M,z \vert M_{cl}, z_f) \over dz} &&dM
=  f_q \, {d f(M,\delta_c (z) \vert M_{cl}, \delta_c(zf)) \over dz} dM
\nonumber\\
&& + f_q f(M,\delta_c (z) \vert M_{cl}, \delta_c(zf)) dM
\nonumber\\
&& \times
\int_{2 M}^{M_{cl}} {d^2 p \over dM' dz}
(M \rightarrow M' \vert z) dM' \,,
\label{eq:rate}
\eea
with the condition that ${d n_q \over dz} =0$, for $z < z_n$,
where ${dn_q \over dz}\vert_{ z<z_n} >0$.
The integral on the right hand side is evaluated using eqn(\ref{eq:lc}).
Here we have also changed the upper limit of the integration to $M_{cl}$.
The integrand is a rapidly decreasing function
of $M'$ and the value of the integral depends mostly on the lower limit.

Here, the threshold density contrast in a cosmological constant dominated
universe is given by a fit given by Kitayama \& Suto (1996),
\be
\delta_c(z)= 1.68 [g(z=0)/g(z)] [1+0.0123 \log \Omega_m(z)]\,.
\ee
In our calculations, we have used a fit for $g(z)$ from Carroll, Press
\& Turner (1992),
\bea
&&g(\Omega_m (z), \Omega_{\Lambda} (z)) \nonumber\\
&& \sim { 5 \Omega_m (z)
\over 2 [\Omega_m(z)^{4/7} -\Omega_{\Lambda}(z) +(1+{\Omega_m(z) \over 2})
(1+{\Omega_{\Lambda}(z) \over
70})]} \,,
\eea
where (Lahav \etal 1991),
\bea  
\Omega_m (z)&=&\Omega_{m0} (1+z)^3/[\Omega_{m0} (1+z)^3 +\Omega_{\Lambda 0}]
\,, \nonumber\\
\Omega_{\Lambda}(z)&=& \Omega_{\Lambda 0} /[\Omega_{m0} (1+z)^3 +
\Omega_{\Lambda 0}] \,.
\label{eq:om}
\eea

We emphasize here that the above formalism leads to a {\it conservative}
estimate of abundance of quasars in a cluster (and, therefore, the final
heat input), because (a) we ignore the increased pace of growth of
perturbation and the merging rate inside a cluster,
and (b) the lower limit of the integration could in reality be smaller
than $2 M$, which is probably the reason the rate of formation still
turns negative at low redshifts even after the addition of merger term.

To summarise the work in this section, we have used the existing ideas
for relating the quasar formation rate to the Press-Schechter
mass function, to estimate the rate of formation of quasars {\it inside 
clusters} (as a function of cluster mass and formation redshift), utilizing
the extensions of PS formalism. First, we use eqn (\ref{eq:fq}) to relate
the PS mass function to quasar abundance. The standard PS mass function
(eqn \ref{eq:ps1}) is then replaced by its extension (eqn \ref{eq:ps3}).
Furthermore, we add the contribution due to merger (into larger objects)
(eqns \ref{eq:merg1} \& \ref{eq:int}), again using the extensions of
PS mass function. We finally have the rate of formation of quasars
inside clusters as given by eqn (\ref{eq:rate}).

We use this formalism to calculate the total energy input from outflows
from quasars in a cluster, or, a lower limit to it. 
We next discuss the energy input from individual
outflows, which we will combine with our calculation of formation rate
of quasars for the final result.

\section{Work done by quasar outflows}

In this section we calculate the energy input from quasar outflows 
into the ambient medium. We model the outflows as they evolve
in the ambient medium and calculate the pdV work done by the outflows.
To begin with, we discuss different kinds of outflows that we consider
and the characteristics of the hosts of quasars.

\subsection{Quasar outflows}

We consider two major types of quasar outflows. For
radio-loud quasars (RLQ), the outflow is in the form of a tightly
collimated jet, which deposits energetic  particles into a cocoon
which expands against the surrounding medium. These outflows are
characterised by the kinetic luminosity of the jet, $L_k$. According
to Willot \etal (1999), this is correlated with the  bolometric
luminosity $L_{bol}$ of the quasar, and that $0.05 \la L_k/L_{bol} \la 1.0$.
We follow FL01 in arguing that since $L_{bol} \sim 10 L_B$ (Elvis \etal (1994)),
the rest-frame B-band luminosity, $L_k \sim L_B$. 

Radio-loud quasars, however, constitute only about $10 \%$ of the
total population of quasars (Stern \etal 2000). We therefore define
a factor $f_o$ for the fraction of quasars with  outflows, and define
$f_o \sim 0.1$ for our RLQ model. The fraction $f_o$ here denotes the
number of radio relics/lobes per halo, since a radio loud quasar may
have several outbursts of radio activity.
This fraction is therefore a very conservative estimate since it is obtained
from observed radio luminosity function and does not take into account
the existence of radio relics in clusters.

Another important kind of outflows
are encountered in broad absorption line (BAL) quasars. The absorption
troughs are thought to be due to absorbing clouds flowing out of
the quasars with velocities up to $0.1 c$. Although they are
encountered in about $10 \%$ of the quasars, it is believed that
all quasars have such outflows (all the time) and the covering fraction of the
BAL outflows is about $10 \%$ (Weymann \etal 1991; Weymann 1997). Some authors
also believe that BAL outflows have a limited lifetime (especially the low
ionization BALs)
and that they have a large covering fraction in
the early phase of a quasar (Voit \etal 1993). For our calculation,
we use a fraction $f_o \sim 1$ for the BAL outflows. We discuss the
effect of the uncertainty in these factors on the final result in \S 5.

We model the BAL outflows as having a kinetic luminosity $L_k$. Following
FL01, if $N_H$
is the column density of the absorbing gas, $f_c$ the covering fraction,
and $R_{BAL}$ is the size of the absorption system, then $L_k$ is
related to the outflow velocity $v_{BAL}$ as $L_k  \sim 2 \pi f_c N_H
m_p R_{BAL} v_{BAL}^3$. The observed range of these parameters are as
follows: $v_{BAL} \la 0.1 c$, $f_c \sim 0.1$ (Weymann 1997; but see above),
$R_{BAL} \sim 1\hbox{--}500 $pc and $N_H \sim 10^{22}\hbox{--} 10^{23}$ 
cm$^{-2}$ (Krolik 1999; Gallagher \etal 1999). 
For these values, the magnitude of $L_k$ is close to that
of $L_B$. FL01 also argued that for BAL winds $L_k \sim 0.1\hbox{--}100
L_B$, and finally assumed $L_k \sim L_B$. Since this estimate depends
crucially on a number of uncertain parameters (for example, the fact that
the absorption column density in optical measurement is much smaller
than the above mentioned X-ray column density), it may not really be a 
conservative estimate, but it does provide a simple scaling which we
hope is not too unreasonable. The estimate $L_K \sim L_B \sim 0.1 L_{Edd}$
is probably not a conservative estimate, but an upper limit, 
in that for a covering fraction
of $10 \%$, the mechanical luminosity of the wind could not be larger than
$0.1$ of the Eddington rate. Keeping all these uncertainties in mind,
we assume that $L_k \sim L_B$ for BAL outflows.

We then need to connect $L_B$ of a given quasar with the properties of its
halo. Firstly, as Haiman \& Loeb (1998) have shown, the mass of the
black hole at the centre is related to $L_B$, as,
\be
M_{BH}={1 \over 0.093} \Bigl ( {L_B \over 1.4 \times 10^{38} \, {\rm erg}
\, {\rm s}^{-1} }\Bigr ) \, M_{\odot} \,
\label{eq:bh}
\ee
assuming that the quasar radiates at the Eddington luminosity. The factor
of $0.093$ reflects the fraction of the Eddington luminosity radiated in
the B-band, taken from the median quasar spectrum of Elvis \etal (1993).
Statistically speaking, 
we therefore assume that a fraction
$f_q$ (eqn(\ref{eq:fq})of all black holes radiate at 
the Eddington rate for a life time of
$\sim 10^7$ yr, while the rest does not radiate at all.

Secondly, the correlation between the central black hole mass and the
total baryonic mass of the galaxy (Magorrian \etal 1998; Gebhardt \etal 2000)
gives $M_{BH} \sim 4 \times 10^{-4} M_h$, where $M_h$ is the total mass
of the galaxy, using a value of $M_{BH}/M_{baryonic} \sim 2\hbox{--}3 \times
10^{-3}$, and $M_{baryonic}/M_h \sim \Omega_b /\Omega_0 \sim 0.2$.

As far as the collimation is concerned, the geometry of  BAL outflows
is still uncertain, whereas the radio jets are well collimated. Since
some models do suggest a modest collimation even in BAL outflows, with
a covering fraction of $f_c \sim 0.1$ (Weymann  1997), we use the
idea of collimated outflows for outflows from both radio-loud and BAL quasars.

We next discuss the evolution of the outflows from radio-loud quasars and
calculate
the fraction of its total kinetic luminosity that it deposits into the
surrounding medium in the form of pdV work. For concreteness,
we will assume that the
corresponding fraction for BAL outflows has similar values, and keep in
mind the uncertainty in the geometry and energetics of BAL outflows.

\subsection{Evolution of outflows}

The standard scenario for outflows from radio loud quasars involves a `cocoon'
surrounding the core and the jet, and consisting of a shocked ambient medium
and shocked jet material (Scheuer 1974; Blandford and Rees 1974). Begelman \&
Cioffi (1989) constructed a simple model of the evolution of a cocoon
in which the cocoon is overpressured against the ICM.
In their model, the expansion along the jet axis is determined by
the balance of the thrust of the jet and the ram pressure, whereas the thermal
pressure of the cocoon drives along the direction perpendicular to the jet
axis. Results of numerical simulations agree with this
scenario (Loken et al. 1992; Cioffi \& Blondin 1992).

Here we adopt the model of the evolution of cocoons following the
approach of Bicknell \etal (1997), which is based on the Begalman \& Cioffi
(1989) model but includes the pdV work done by the cocoon, 
in order to find the fraction of total energy lost
by the quasar to the ICM through mechanical work (pdV work). Bicknell \etal (1997)
derived this fraction ($f_{pdV}$) to be $f_{pdV}=3/8$, for a homogeneous
ambient medium (their equation (2.13)). 

We will, however, calculate this fraction from numerical solution
of the equations governing the evolution of the cocoon, for the following
reasons. Firstly, the derivation mentioned above implicitly assumes that
the mean pressure averaged over the hotspot region is equal to the mean
lobe pressure (in the language of Bicknell \etal (1997), this means
$\zeta \sim 1$). In fact, their equation (2.13) shows 
that for constant $\zeta$, one has in general,
$P_c {dV_c \over dt} = (1+2 \zeta) L_j/8 \zeta$, which recovers the fraction
$3/8$ for $\zeta=1$. In reality, however, 
this ratio does not remain a constant in time. 
Secondly, this derivation is valid only during the period when the jet
is active. Even after the jet switches off, the cocoon, however, continues
to evolve as a result of its overpressure until it reaches an equilibrium
pressure with the ambient medium (see also Nath 1995). The cocoon, therefore,
continues to do pdV work even after the jet switches off, and the inclusion
of this process will lead to an upward revision of the fraction of total
energy that is lost in pdV work. Besides, there seems to be some confusion
in the literature regarding the fraction. For example, Inoue \& Sasaki (2001)
have recently adopted a fraction $f_{pdV}=1/4$ in their calculation of
energy input into the surrounding gas. 

We therefore calculate this fraction by numerically solving the equations
of cocoon evolution.


We consider two collimated steady jets advancing into the
ambient ICM. The thermalized jet matter and the shock-compressed ICM matter
form a cocoon around the jets and the cocoon expands with shocks advancing in
directions both parallel and perpendicular to the jet axis. After this stage
of evolution, when the jet turns off after a
lifetime of $t_{life} \sim 3\times10^{7}$ years (Kaiser 2000), 
cocoons still retain high
pressure. They cool radiatively and expand due to its 
overpressure till it reaches a pressure equilibrium with the ambient medium.
Thus the relevant equations are:

\begin{eqnarray}
{dr_h \over dt}&=& \Bigl ({L_j \over A_h\rho_a\beta c}\Bigr )
^{1/2} ,\qquad t<t_{life}\nonumber\\
&=& \Bigl ({P_c\over \rho_a}\Bigr )^{1/2} , \qquad t>t_{life}\\
{dr_c \over dt}&=& \Bigl ({P_c \over \rho_a}\Bigr )^{1/2} , \\
{dE_c \over dt}&=& L_j - P_c{dV_c \over dt} 
\end{eqnarray}
where $L_{j}$ is the jet luminosity, 
$\rho_{a}$ is the density of the ambient
medium and $\beta c$ is the velocity of the jet material.
As the jet is highly relativistic, $\beta \sim 1$. The averaged hotspot area
$A_h \sim 30$ kpc$^2$ (Bicknell \etal 1997) is 
assumed to be larger than the radius of the jet, according to the
`dentist's drill' model of the jet (Scheuer 1982).
Here $r_h$ is the length of the jet or the distance of the hotspot from the
centre of the galaxy, $r_c$ is the half-width of the cocoon at the centre
and $V_c$ is the volume of the cocoon given by $V_c = \epsilon_v(2\pi r_c^2)
r_h$ where $\epsilon_v$ is the geometrical factor depending on the shape
of the cocoon. For our calculations we have taken the shape of the cocoon as
biconical and so $\epsilon_v \sim 1/3$.
Finally $P_c$ is the pressure inside the cocoon given by $P_c=(E_c/V_c)
(\gamma - 1)$ where $\gamma = 4/3$ and $E_c$ is the total energy inside the
cocoon given by $E_c = L_jt$ till $t<t_{life}$ and $E_c = L_jt_{life}$ 
afterwards, where $t_{life}$ is the lifetime of the jet.

This is admittedly a  simplified model of the evolution of the cocoon.
In reality, after the jet switches off, one expects Rayleigh-Taylor and
Kelvin-Helmholtz instabilities to distort the cocoon, giving rise to
`buoyant' plumes. This phase of the evolution of the cocoon, and its
effect on the ambient medium, has been recently addressed by various authors
(Gull \& Northover 1973; Churazov \etal 2000; Br\"uggen \& Kaiser
2000), mainly with the help of numerical simulations. It is possible that
this phase adds substantial heating to the intracluster medium, but short
of doing a numerical simulation it is difficult to assess its importance.
We have also neglected the loss of energy through radiation, as 
modeling such losses would involve detail knowledge of various parameters
(e.g., electron energy spectrum and the possibilities of re-acceleration
of electrons). With the uncertainties involved in modeling these process,
it seems reasonable to adopt the above simplified picture as a pointer
and keep the uncertainties in mind while discussing the final result.
In light of this discussion, we will also calculate the final result
with a value of $f_{pdV}=3/8$ as in Bicknell \etal (1997), which we
will adopt as a conservative lower limit.

\begin{figure}
\begin{center}
{\vskip-4mm}
\psfig{file=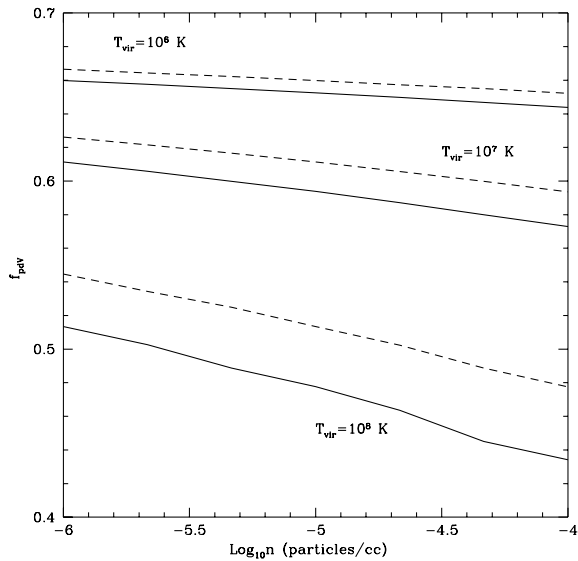, width=8cm}
\end{center}
{\vskip-3mm}
\caption{
The dependence of $f_{pdV}$ with the ambient density $n$
is shown for various ambient temperatures $T$ for various  
$L_k$ (erg/s) with and without radiation loss. Solid and dashed lines are 
for $L_k=10^{46}$ erg/s and $10^{47}$ erg/s respectively without radiation
loss. Dotted lines are for $L_k=10^{46}$ with radiation loss.
}
\end{figure}

We numerically calculate the volume of the cocoon as it
grows into the ICM and the pressure inside the cocoon at each step,
and add up the pdV work to get the final amount of energy lost in this mode.
The evolution of the cocoon is continued until
 the pressure inside the cocoon becomes equal to the
ambient pressure, $n_a k_B T_a$ ($n_a=\rho_a/\mu m_p$), where
$T_a$ is the temperature of
the ICM. The fraction $f_{pdV}$ is calculated by taking the ratio of the
total energy lost through mechanical work to the total energy (that is
$L_jt_{life}$). (We found that for the relevant values of the ambient
medium parameters, the time scale to reach pressure equilibrium is always
larger than $t_{life}$.) The dependence of $f_{pdV}$ on the ambient
density, 
is shown in Figure 2 for $T_a=10^6, 10^7$ and $10^8$ K, and 
for $L_k=10^{46}$ (solid lines) and $10^{47}$ erg/s (dashed lines).

The plot shows that the fraction $f_{pdV}$ is a function of temperature of the
cluster and also the density of the ambient medium. The general trend
is that the
fraction reduces at higher temperatures and higher densities.
This is because of the fact that the cocoon reaches pressure equilibrium
with the ambient medium sooner for a higher pressure environment (higher $T_a$
and $n_a$), and the total pdV work ends up being smaller.
The plot also shows that the fraction $f_{pdV}$ depends weakly on the jet
luminosity for lower temperatures ($T_a \le 10^7$ K), whereas there is
a bit of a difference for $T_a \sim 10^8$ K.



The fraction $f_{pdV}$ calculated above is somewhat larger than that has
been used in the literature, for ambient medium with low pressure. 
The difference is mainly the result of 
our inclusion of cocoon evolution even after the jet has
switched off. Incidentally, Inoue \& Sasaki (2001), while using
a value of $f_{pdV}=1/4$, discussed the possibility that this fraction
could be larger in reality, {\it because} of its continued evolution
after the switching off of the jet (their \S 3.2). 

In our calculation for the total pdV work done by BAL and RLQ outflows,
we will use the values of $f_{pdV}$ obtained above. As mentioned earlier,
the energetics and geometry of BAL outflows are not clear at present. For
concreteness, we have worked out the case of RLQ outflows in detail,
and we will use the same values of $f_{pdV}$ 
for BAL outflows as well.

\section{Heating of the ICM}

Equipped with the knowledge of the rate of formation of quasars
in clusters (equation \ref{eq:rate}) and the fraction of total energy
which is deposited as pdV work by the outflows from them (\S 3.2), we
are now in a position to calculate the total amount of non-gravitational
energy provided by quasar outflows in a cluster. 
If we denote the gas fraction of the total cluster mass by $f_{gas}$, then
the total number of gas particles is $\sim M_{cl} f_{gas}/m_p$, where $m_p$ is
the proton mass. It is not yet clear if the gas fraction has a universal
value for clusters of all masses and at all redshifts. A few authors
(see, e.g., Schindler (1999)) have found no correlation of the gas fraction
with the cluster mass, whereas others (e.g., Ettori \& Fabian (1999))
find some correlation (with low mass clusters possessing lower gas fraction).
This correlation has been attributed to the excess energy deposition 
(since low mass clusters are more liable to lose gas from heating than
massive clusters) (see, e.g., Bialek \etal (2000)). Here, we are, however,
trying to calculate the magnitude of this very excess energy. It would
not be appropriate to include this correlation {it a priori} in our
calculation. (We show later that including these correlations only increases
our estimate of excess energy input.)
We have, therefore, used a value of $f_{gas}=0.1$ for our calculation.
The total energy per (gas) particle
 deposited into the ICM of a cluster of mass $M_{cl}$
is then given by,
\bea
\epsilon_{pdV}= &&{m_p \over M_{cl} f_{gas}} \int_{z_m}^{0}
\int_{M_l}^{M_u} {d f_{q,cl}(M,z \vert M_{cl}, z_f) \over dM dz} \nonumber\\
&&  {M_{cl} \over M}
dM \, dz \, f_o \, [L_k t_{life} f_{pdV}(n_a, T_a)] \,,
\label{eq:final}
\eea
where $\Bigl [ {d^2 n_q(M,z \vert M_{cl}, z_f) \over dM dz} dM dz \Bigr ]$ 
is calculated
using equation \ref{eq:rate}. The factor 
$f_o \sim 1$ for the BAL outflows, and $f_o \sim 0.1$ for outflows from
RLQs (\S 3). The redshift $z_m$ is the maximum redshift of heat input. We
later show (Figure 7) that the heat input is negligible for $z \ge 5$.
The density and temperature of the ICM
of a cluster of a given mass ($M_{cl}$) and formation redshift ($z_f$)
 is calculated using (Eke \etal 1998),
\be
T_a=1.65 \times 10^7 (1+z) \Bigl [ {M_{cl} \over 10^{15} h^{-1} \, M_{\odot}}
\Bigr ]^{2/3} \Bigl [ { \Omega_0 \Delta(\Omega_0,z) \over \Omega (z)}
 \Bigr ]^{1/3} \,,
\ee
and,
\be
n_a={M_{cl} f_{gas} \over m_p (4/3) \pi r_{vir}^3} \,,
\ee
where 
\be
\Delta (z)= 18 \pi ^2 + 82  x - 39 x^2 \,
\ee
and $x=\Omega(z)-1$ (Bryan \& Norman 1998), where we use equation \ref{eq:om}
to compute $\Omega (z)$. 
For $r_{vir}$, we use,
\be
r_{vir}=\Bigl ( { 3 M_{vir} \over 4 \pi \Delta(z) \rho_{crit}(z)} \Bigr )^{1/3}
\,,
\ee
where we have used $M_{vir}=M_{cl}$, and $\rho_{crit}$ is the critical
density of the universe. The densities used in the following calculations
range between $10^{-4}\hbox{--}10^{-6}$ cm$^{-3}$.

The integral in equation (\ref{eq:final})  is evaluated using $M_l=10^{12}$
M$_{\odot}$ and $M_u=10^{13}$ M$_{\odot}$, for $z_f=0, 0.5, 1$ for different
values of $M_{cl}$. We present the results for the total non-gravitational
energy input per particle as a function of cluster mass (or, equivalently,
gas temperature) in Figure 4 (for $z_f=0$). The solid curve shows the
heat input calculated using $f_{pdV}$ from \S 3.2. The
dotted line shows the case for a constant $f_{pdv}=3/8$ (Bicknell \etal 1997). 
We show the
results for different $z_f$ in Figure 5 
(against $T$) and Figure 6 (against cluster mass). 

\begin{figure}
\begin{center}
{\vskip-4mm}
\psfig{file=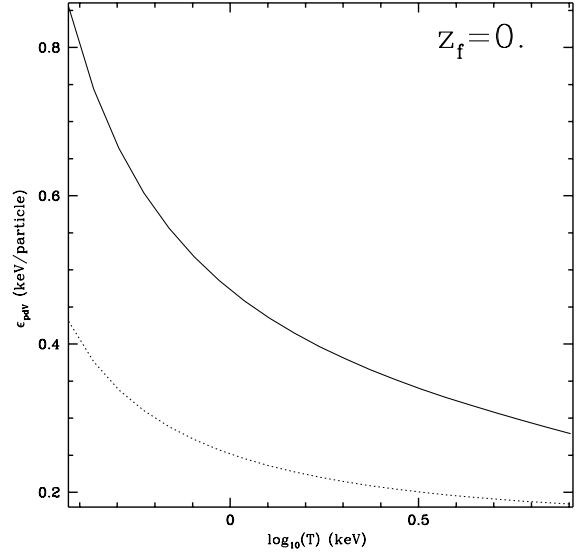, width=8cm}
\end{center}
{\vskip-3mm}
\caption{
Excess energy (in keV) from  BAL outflows is shown as a function of the
cluster virial temperature (keV) for clusters with
$z_f=0$. The solid line shows the result of our calculation using
the density and temperature dependent $f_{pdV}$ and the dotted line
shows the results when $f_{pdV}=3/8$.
}
\end{figure}

We also show in Figure 7
the rate of deposition of energy ($-{d \epsilon_{pdV} \over dz}$)
as functions of the redshift for three clusters of masses 
$M=2 \times 10^{13},
10^{14}$ and $10^{15}$ M$_{\odot}$, all for $z_f=0$. 
All the curves drop
to zero at low redshift because of the condition $dn_{q,cl}/dt=0$ (
in eqn \ref{eq:rate}). In reality the contribution to the heating should
be small but non-zero, and will increase the estimate of excess energy.

\begin{figure}
\begin{center}
{\vskip-4mm}
\psfig{file=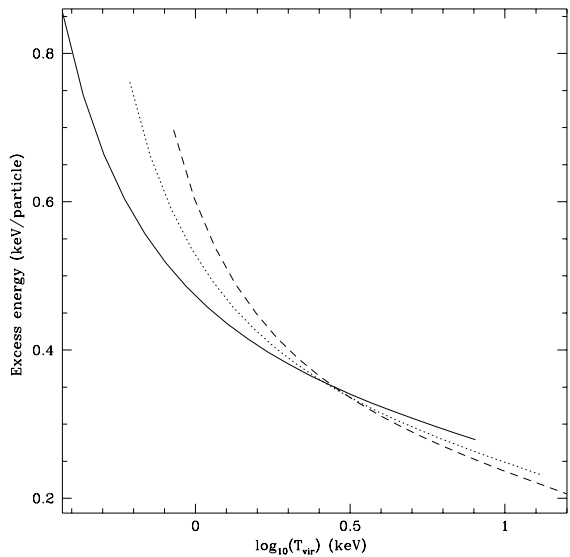, width=8cm}
\end{center}
{\vskip-3mm}
\caption{
The excess energy is shown for $z_f=0$ (solid) $0.5$ (dotted) and $1$ (dashed).
}
\end{figure}

\begin{figure}
\begin{center}
{\vskip-4mm}
\psfig{file=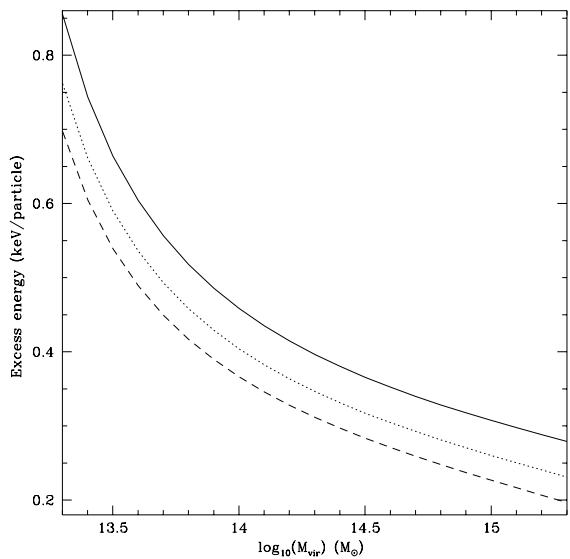, width=8cm}
\end{center}
{\vskip-3mm}
\caption{
The excess energy against the cluster/group mass is shown for 
$z_f=0$ (solid) 
$0.5$ (dotted) and $1$ (dashed).
}
\end{figure}

\section{Discussion}

It has been estimated that the amount of excess energy required to
explain the observation is of order $0.5\hbox{--}3$ keV per particle
(Navarro et al. 1995;
Cavaliere et al. 1997; Balogh et al.1999; Wu et al. 2000). Recently,
however, Lloyd-Davies \etal (2000) have shown from observations of
groups of clusters that an excess energy of $0.44 \pm 0.3$ keV per particle
suffices to explain the excess entropy in groups. They showed that
this can explain the entropy floor for galaxy groups with temperature
$T \la 4$ keV. Borgani \etal (2001) have also shown with the aid of
numerical simulations that excess energy of order $\sim 1$ keV per
particle  reproduces the observations. 

The solid and dashed curves in Figures 4 \& 5 show that the excess energy from
pdV work done by quasar outflows fall in this required
range. It is seen
that the excess energy per particle is larger for clusters or groups
with lower temperature. This is due to two factors: (a) the number
of quasars per unit mass is larger for smaller clusters,
and (b) the fraction of total energy in outflows that is lost in pdV
work is larger for them. We show the results in the case of
a constant $f_{pdV}=3/8$ (as in Bicknell \etal (1997)) with dotted
lines. 
It is interesting to note that even in this case the excess energy is in the
required range ($0.44 \pm 0.3$ keV per particle), especially for groups
with low temperatures, as advocated by Lloyd-Davies \etal (2000).
Incidentally, this is  larger than the estimate of 
excess energy from galactic winds ($\la 0.1$ keV per particle, Wu
\etal (2000)). 

We found that our results for the excess energy 
(solid line in Figure 4) can be approximated
by a fit of type (in keV per particle),
\bea
\epsilon_{pdV} &&\sim 0.258 \Bigl ({T \over 10 \, {\rm keV}} \Bigr )^{-0.193}
+0.033 \Bigl ( {T \over 2 \,
{\rm keV}} \Bigr )^{-1.2}, \, \nonumber\\
&& 0.50 \le T \le 8.0 \, {\rm keV}
\eea
We also found that the results for the excess energy taking 
$f_{pdV}=3/8$ (Bicknell \etal (1997))~(dotted line in Figure 4) can be 
approximated by a fit of type (in keV per particle),
\be
\epsilon_{pdV} \sim 
0.17 +0.045 \Bigl ( {T \over 2 \,
{\rm keV}} \Bigr )^{-0.95} \,, 
0.50 \le T \le 8.0 \, {\rm keV}
\ee

We compare our results with the data from Lloyd-Davies \etal (200) in Figure 8.
We show the predictions of our calculation as the thin solid curve
(corresponding to the solid curve in Figure 4),
where the data points have been taken from that of Lloyd-Davies \etal (2000)
(their Figure 9). We also show the result of the calculations for
$f_{pdV}=3/8$ by the thick solid line.
The data points refer to the binding energy of a (constant)
central fraction ($0.004$) of the virial mass of groups 
and clusters. The dashed line shows the case for self-similar
models, of type $E \propto T$, derived from the data points for rich
clusters. The dotted line shows their fit (with a constant
excess energy of $0.44$ keV per particle) along with a formal 1 $\sigma$
confidence interval shown by the shaded region.
The figure shows that our predictions are consistent with
the data available at present.
The thick line (corresponding to $f_{pdV}=3/8$) falls close to the fit
provided by Lloyd-Davies \etal (2000), whereas the thin line (using
$f_{pdV}$ from Figure 3) somewhat overestimates the heat input at the low mass
end. The thick line can be viewed as a conservative estimate of the heat
input, since it uses $f_{pdV}=3/8$. The thin line, however, provides
an estimate of the heat input if $f_{pdV}$ is much larger than $3/8$. We
should remind ourselves here that we have calculated $f_{pdV}$ for 
radio galaxies and used the same values for BAL outflows. If a more accurate
estimate of $f_{pdV}$ for BAL outflows is worked out in the future, the
resulting heat input into the ICM could then be scaled accordingly using
Figure 8.


We would like to emphasize here again that the our calculation 
provides a conservative estimate of the excess energy, for reasons outlined
in \S 2. Moreover, we have used a constant density and temperature in time
for the ICM gas (for clusters with a given $z_f$), which is not very 
realistic. In reality, the density at higher redshift will be smaller,
and the inclusion of a density dependent $f_{pdV}$ will only increase
the estimate of excess energy (since this fraction increases with decreasing
density). 

The curves for clusters forming at different epochs
show  excess energies to decrease somewhat for clusters with higher formation
redshift. 

\begin{figure}
\begin{center}
{\vskip-4mm}
\psfig{file=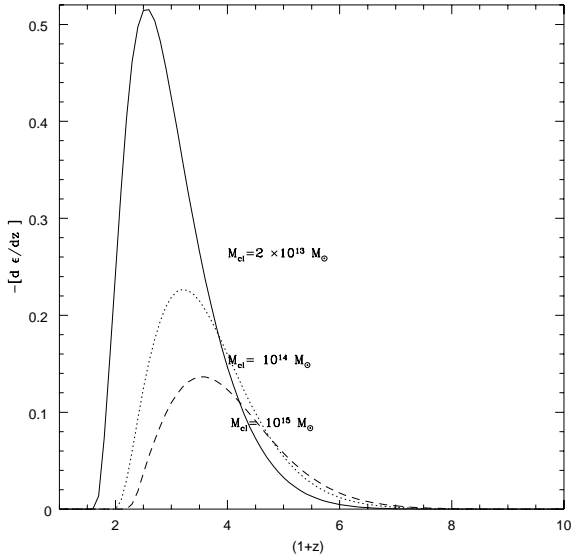, width=8cm}
\end{center}
{\vskip-3mm}
\caption{
The rate of deposition of excess energy ($-d \epsilon_{pdV} /dz$) is shown
as a function of redshift, for $M_{cl}=2 \times 10^{13}$ M$_{\odot}$
(solid line), $10^{14}$ M$_{\odot}$ (dotted line) and 
$10^{15}$ M$_{\odot}$ (dashed line)
}
\end{figure}

The curves of Figures 4 \& 5 
assume $f_o \sim 1$, which is relevant for BAL 
outflows. The excess energy from RLQ outflows will be one tenth of these
curves, showing the difficulty of using radio galaxies as the only source
of non-gravitational heating, if conservative estimates for their
kinetic luminosities are used.
Recently, Inoue \& Sasaki (2001) have used the radio luminosity
functions of Willot \etal (2001) and Ledlow \& Owen (1996) to determine
the abundance of radio galaxies in clusters, and finally to estimate the
total pdV work done by the cocoons of these radio galaxies. They estimated
an excess energy of order $1$ keV per particle for rich clusters like the
Coma cluster, and 
also for poor groups, assuming that their ratio of radio galaxies per unit
cluster mass is universal. From our calculation, we find an excess energy
from only radio-loud quasars that is an order of magnitude {\it smaller} than
their estimate. 
It is possible that the assumptions leading to the estimate of $L_k$
are at the source of this difference (see, e.g., the discussion on the
uncertainty in the factor $f_j$ in their \S 3.2).

\begin{figure*}
\begin{minipage}{140mm}
\begin{center}
{\vskip-4mm}
\psfig{file=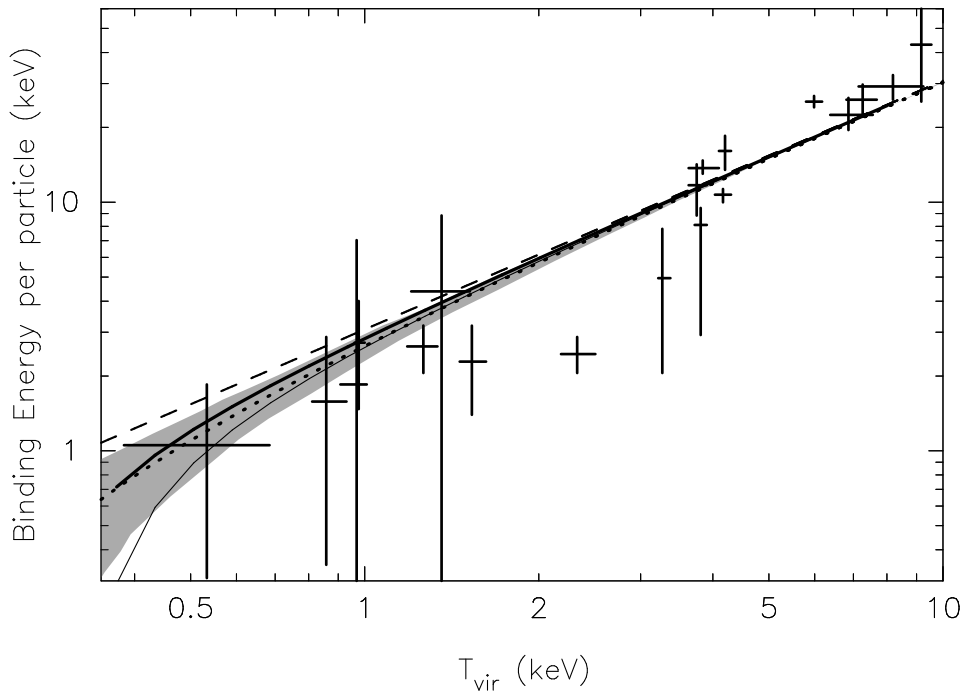, angle=270, width=16cm}
\end{center}
{\vskip-3mm}
\caption{
The prediction from our calculations is presented in the form of the
final
binding energy per particle of the central region of groups and clusters
against the gas temperature. The data points are from the Figure 9 of
 Lloyd-Davies
\etal (2000), and the dashed line refers to their fit $E \propto T$,
derived from the data points for clusters with $T \ge 4$ keV. The dotted
line refers to their second fit, with a constant excess energy of
$0.44$ keV per particle (subtracted from the binding energy)along with a 
formal 1 $\sigma$ confidence interval shown by the shaded region. 
The thick solid line uses $f_{pdV}=3/8$ and the thin solid line uses
$f_{pdV}$ from Figure 3.
}
\end{minipage}
\end{figure*}

The evolution of the rate of energy input (Figure 7) shows that
the heat input before $z \sim 5$ is almost negligible. 
Most of the heating occurs in the range $z
\sim 1\hbox{--}4$.  It is also clearly seen that the ICM of poor groups
is heated at lower redshifts, compared to the gas in massive clusters.
This follows from the simple consideration that the evolution of objects
in a given mass range (here, that of the quasars) typically occurs earlier 
in more massive clusters. Our results, therefore, suggests that the ICM
in clusters were ``preheated'', before the major mergers took
place in them, whereas, the ICM in groups of galaxies were heated at
epochs similar to that of their formation.
It is interesting to note that the energy input epoch
lies below the upper limit from recent observations of Lloyd-Davies \etal 
(2000). It is also consistent with the limit on the range of redshifts as
shown by Yamada \& Fujita (2001) given the level of uncertainty.

Recently, accelerated particles from shocks as a result of the formation of 
clusters have been hypothesized to be the source for the diffuse
gamma ray background (Loeb \& Waxman 2000). This gamma ray production will
be suppressed, however, if the gas in the clusters were heated substantially
at earlier epochs (Totani \& Inoue 2001). Our results show that the ICM
in massive clusters were preheated (and earlier than the ICM in groups),
and the suppression of the gamma ray production will, therefore, be an
important effect, if confirmed.

Finally, we discuss the uncertainties involved in our calculation. Apart
from the uncertainties in cosmological parameters, the major uncertainties
lies in the factors $f_q$ (\S 2) (connecting the abundances of quasars
with PS mass function) , $f_o$ (the fraction of quasars with outflows)
,$f_{pdV}$ and $f_{gas}$. 
Among these, the most uncertain factor is $f_o$, which we
have assumed to be of the order of unity for BAL outflows. The uncertainty
in this factor will be reflected in the uncertainty of the final heat input
(with a direct proportionality; see eqn \ref{eq:final}). The uncertainty
in $f_{pdV}$ has been already discussed earlier, and we found that even if
$f_{pdV}$ is as low as $3/8$ for all cases, the final excess energy is
certainly larger than that from supernovae driven winds, and is still
within the required range of excess energy, especially for loose groups.
As far as the uncertainty in $f_{gas}$ is concerned,
we have also done our calculation with  a varying $f_{gas}$, e.g.,
of the type,
\be
f_{gas}=0.15 (1+z)^{-0.5} \, (M_{cl}/10^{15} h^{-1})^{0.1} \,,
\ee
as has been advocated by Ettori \& Fabian (1999), and we have found that
the excess energy is approximately doubled  in this case. It is, however,
not clear if this correlation is a result of the excess energy, and, so,
it would not be appropriate to attach much significance to this result.
We have also varied the lower limit in our estimate of $f_q$ (eq. 
\ref{eq:fq})
and found that changing the lower limit from $10^{12}$ M$_{\odot}$ to 
$10^{11}$ M$_{\odot}$ increases the final heat input by only  $\sim 10$ \%.
This is because of the fact that the increase in the number of quasars
is compensated by the decrease in their mechanical luminosity.
Lastly, we have already discussed in detail the uncertainty in the
net formation rate of quasars in clusters, and as explained in \S 2, our
approach here has been very conservative, and the final results should
be regarded as conservative estimates in this regard.

\section{Summary}

We have calculated the excess energy deposited by quasar outflows 
in clusters in order to explain the observations of excess
entropy in groups and clusters of galaxies. We summarise our findings
below:

(1) We have used the extended Press-Sechter formalism to derive a formation
rate of quasars inside clusters and groups, as a function of the
cluster/group mass and its formation redshift. 

(2) We have calculated the fraction of the kinetic luminosity of outflows
(RLQ and BAL outflows) that is deposited onto the ambient
medium, as a function of the density and temperature of the ambient medium.
For outflows from radio-loud quasars,
we have included the evolution of the cocoon after the jet turns off.

(3) The final excess energy from the mechanical 
work done by quasar outflows 
is found to be of order $0.18\hbox{--}0.85$ keV per particle,
and is consistent with the data available at present.
The excess energy in this scenario comes mainly from BAL outflows, with
radio galaxies supplying about a tenth of the total. 
Keeping in mind the uncertainties in the estimate of energetics and abundances
of radio and BAL outflows, we conclude that both radio galaxies and BAL
outflows are promising candidates for heating the ICM.
We found that this excess energy increases with decreasing 
mass of the cluster/group.  This prediction could be tested with better
data in the near future. 
The excess energy does not depend strongly on the formation redshift.

(4) The epoch of heating is found to be in the range
 $z \sim 1\hbox{--}4$, where this epoch
is at lower redshifts for low mass clusters.

\bigskip\bigskip
The authors would like to thank Nahum Arav, Andrew Blain, Sergio
Colafrancesco, Zoltan Haiman, Subhbrata Majumdar
and Trevor Ponman for valuable discussions. We have also benefitted from
the detail comments of the anonymous referee.
We are grateful to Ed
Lloyd-Davies for supplying the data points for binding energy.
We would also like to thank
Rekhesh Mohan for help with plotting packages.


\begin{thebibliography}{}

\bibitem[\protect\citename{bn}%
]{bchokshi91}
Bahcall, J. N. \& Chokshi, A. 1991, ApJ, 380, L9

\bibitem[\protect\citename{bn}%
]{balogh99}
Balogh, M. L., Babul, A. \& Patton, D. R. 1999, MNRAS, 307, 463

\bibitem[\protect\citename{bn}%
]{bc89}
Begelman, M. C. \& Cioffi, D. F. 1989, ApJ, 345, L21

\bibitem[\protect\citename{bn}%
]{bem00}
Bialek, J. J., Evrard, A. E. \& Mohr, J. 2001, ApJ, 555, 597

\bibitem[\protect\citename{bn}%
]{bick97}
Bicknell,  G. V., Dopita, M. A. \& O'Dea, C. P. O. 1997, ApJ, 485, 112

\bibitem[\protect\citename{bn}%
]{bl93}
Blain, A. W. \& Longair, M. S. 1993, MNRAS, 265, L21


\bibitem[\protect\citename{bn}%
]{borgani01}
Borgani, S., Governato, F., Wadsley, J., Menci, N., Tozzi, P., Lake, G.,
Quinn, T. \& Stadel, J. 2001, ApJL, 559, L71

\bibitem[\protect\citename{bn}%
]{bow91}
Bower, R. G. 1991, MNRAS, 248, 332


\bibitem[\protect\citename{bn}%
]{bk01}
Br\"uggen, M. \& Kaiser, C. R. 2001, MNRAS, 325, 676


\bibitem[\protect\citename{bn}%
]{bn98}
Bryan, G. L. \& Norman, M. L. 1998, ApJ, 495, 80

\bibitem[\protect\citename{bn}%
]{c97}
Cavaliere, A., Menci, N. \& Tozzi, P. 1997, ApJ, 484, L21

\bibitem[\protect\citename{bn}%
]{co00}
Chiu, W. A. \& Ostriker, J. P. 2000, ApJ, 534, 507

\bibitem[\protect\citename{bn}%
]{cbkbf00}
Churazov, E., Br\"uggen, M., Kaiser, C. R., B\"ohringer, H.,
Forman, W. 2001, ApJ, 554, 261

\bibitem[\protect\citename{bn}%
]{dfj91}
David, L. P., Forman, W. \& Jones, C. 1991, ApJ, 380, 39

\bibitem[\protect\citename{bn}%
]{daly94}
Daly, R. A. 1994, ApJ, 426, 38

\bibitem[\protect\citename{bn}%
]{Eke98}
Eke, V., Navarro, J. F. \& Frenk, C. S. 1998, ApJ, 503, 569

\bibitem[\protect\citename{bn}%
]{Ellingson91}
Ellingson, E., Green R. F. \& Yee, H. K. C. 1991, ApJ, 371, 49

\bibitem[\protect\citename{bn}%
]{elvis94}
Elvis, M. \etal 1994, ApJS, 95, 1

\bibitem[\protect\citename{bn}%
]{ef99}
Ettori, S. \& Fabian, A. C. 1999, MNRAS, 305, 834

\bibitem[\protect\citename{bn}%
]{ev91}
Evrard, A. E. \& Henry, J. P. 1991, ApJ, 383, 95

\bibitem[\protect\citename{bn}%
]{F96}
Fisher, K. B., Bahcall, J. N., Kirhakos, S., Schneider, D. P. 1996,
ApJ, 468, 469

\bibitem[\protect\citename{bn}%
]{fl01}
Furlanetto, S. \& Loeb, A. 2001, 556, 619 (FL01)

\bibitem[\protect\citename{bn}%
]{g99}
Gallagher, S. C. \etal 1999, ApJ, 519, 549

\bibitem[\protect\citename{bn}%
]{ge00}
Gebhardt, K. \etal 2000, ApJ, 543, 5

\bibitem[\protect\citename{bn}%
]{gn73}
Gull, S. F. \& Northover, K. J. E. 1973, Nature, 224, 80


\bibitem[\protect\citename{bn}%
]{hl98}
Haiman, Z. \* Loeb, A. 1998, ApJ,  503, 505


\bibitem[\protect\citename{bn}%
]{is01}
Inoue, S. \& Sasaki, S. 2001, ApJ, 562, 618 

\bibitem[\protect\citename{bn}%
]{hr93}
Haehlent, M. \& Rees, M. J. 1993, MNRAS, 263, 168

\bibitem[\protect\citename{bn}%
]{k91}
Kaiser, N. 1991, ApJ, 383, 104

\bibitem[\protect\citename{bn}%
]{ka99}
Kaiser, C. R. \& Alexander, P. 1999, MNRAS, 302, 515

\bibitem[\protect\citename{bn}%
]{k00}
Kaiser, C. R. 2000, A\&A, 362, 447

\bibitem[\protect\citename{bn}%
]{ks96}
Kitayama, T. \& Suto, Y. 1996, ApJ, 469, 480

\bibitem[\protect\citename{bn}%
]{ky00}
Kravtsov, A. V. \&  Yepes, G. 2000, MNRAS, 318, 227

\bibitem[\protect\citename{bn}%
]{k99}
Krolik, J. 1999, Active Galactic Nuclei (Princeton University Press:
Princeton)


\bibitem[\protect\citename{bn}%
]{lc93}
Lacey, C. \& Cole, S. 1993, MNRAS, 262, 627

\bibitem[\protect\citename{bn}%
]{lahav91}
Lahav, O., Lilje, P. B., Primack, J. R., Rees, M. J. 1991, MNRAS, 251, 128

\bibitem[\protect\citename{bn}%
]{lo96}
Ledlow, M. J. \& Owen, F. N. 1996, AJ, 112, 9

\bibitem[\protect\citename{bn}%
]{lloyd00}
Lloyd-Davies, E. J., Ponman, T. J. \& Cannon, D. B. 2000, MNRAS, 315, 689

\bibitem[\protect\citename{bn}%
]{lw00}
Loeb, A. \& Waxman, E. 2000, Nature, 405, 156

\bibitem[\protect\citename{bn}%
]{mag99}
Magorrian, J. \etal 1998, AJ, 115, 2285

\bibitem[\protect\citename{bn}%
]{me97}
Mohr, J. J. \& Evrard, A. E. 1997, ApJ, 491, 38

\bibitem[\protect\citename{bn}%
]{n95}
Nath, B. B. 1995, MNRAS, 274, 208

\bibitem[\protect\citename{bn}%
]{nfw95}
Navarro, J. F., Frenk, C. S. \& White, S. D. M. 1995, MNRAS, 275, 720


\bibitem[\protect\citename{bn}%
]{pm99}
Percival, W. \& Miller, L. 1999, MNRAS, 309, 823

\bibitem[\protect\citename{bn}%
]{po99}
Ponman, T. J., Cannon, D, B. \& Navarro, J. F. 1999, Nature, 397, 135


\bibitem[\protect\citename{bn}%
]{s94}
Sasaki, S. 1994, PASJ, 46, 427

\bibitem[\protect\citename{bn}%
]{sch82}
Scheuer, P. A. G. 1982, in IAU Symp. 97, Extragalactic RAdio Sources, ed.
D. S. Heeschen \& C. M Wade (Dordrecht: Reidel), 163

\bibitem[\protect\citename{bn}%
]{schindler99}
Schindler, S. 1999, A\&A, 349, 435

\bibitem[\protect\citename{bn}%
]{stern00}
Stern, D. \etal 2000, AJ, 119, 1526

\bibitem[\protect\citename{bn}%
]{toi01}
Totani, T. \& Inoue, S. 2002, Astroparticle Physics, 17, 79

\bibitem[\protect\citename{bn}%
]{vs99}
Valageas, P. \& Silk, J. 1999, A\&A, 350, 725

\bibitem[\protect\citename{bn}%
]{voit93}
Voit, G. M., Weymann, R. J. \& Korista, K. T. 1993, ApJ, 413, 95

\bibitem[\protect\citename{bn}%
]{wey91}
Weymann, R. J., Morris, S. L., Foltz, C. B. \& Hewett, P. C. 1991,
ApJ, 373, 23


\bibitem[\protect\citename{bn}%
]{wey97}
Weymann, R. J. 1997, in Mass Ejection from AGN, ed. Arav, N., Shlosman, I.
\& Weymann, R. J. (ASP Press: San Fransisco), p.3

\bibitem[\protect\citename{bn}%
]{wil01}
Willot, C. J., Rawlings, S., Blundell, K. M. \& Lacy, M. 1999, MNRAS, 309,
1017

\bibitem[\protect\citename{bn}%
]{wold01}
Wold, M., Lacy, M., Lilje, P. B., Serjeant, S., 2001, MNRAS, 323, 231

\bibitem[\protect\citename{bn}%
]{wfn99}
Wu, K. K. S., Fabian, A. \& Nulsen, P. E. J. 2000, MNRAS, 318, 889

\bibitem[\protect\citename{bn}%
]{yss99}
Yamada, M., Sugiyama, N. \& Silk, J. 1999, ApJ, 622, 66

\bibitem[\protect\citename{bn}%
]{yf01}
Yamada, M. \& Fujita, Y. 2001, ApJ, 553, L145


\end{thebibliography}
\end{document}